\documentclass[preprint,superscriptaddress]{revtex4-2}
\usepackage[cp1251]{inputenc} 
\usepackage{amsfonts,amssymb,amsmath,amsthm,graphicx}

\begin{document}

\title{Casimir-Polder interaction with Chern-Simons boundary layers}

\author{Valery N. Marachevsky}
\email{maraval@mail.ru; v.marachevsky@spbu.ru}
\author{Arseny A. Sidelnikov}
\email{st074065@student.spbu.ru}

\affiliation{ St. Petersburg State University, 7/9 Universitetskaya
nab., St. Petersburg 199034, Russia}

\begin{abstract}
Green functions scattering method is generalized to consider mixing
of electromagnetic polarizations after reflection from the plane
boundary between different media and applied to derivation of the
Casimir-Polder potential in systems with Chern-Simons plane boundary
layers. The method is first applied to derive the Casimir-Polder
potential of an anisotropic atom in the presence of a Chern-Simons
plane boundary layer on a dielectric half-space. Then a general
result for the Casimir-Polder potential of an anisotropic atom
between two dielectric half-spaces with Chern-Simons plane parallel
boundary layers is derived. The Casimir-Polder potential of an
anisotropic atom between two Chern-Simons plane parallel layers in
vacuum is expressed through special functions. Novel P-odd
three-body vacuum effects are discovered and analyzed in the system
of two Chern-Simons plane parallel layers and a neutral atom in its
ground state between the layers. Remarkably, P-odd three-body vacuum
effects arising due to $180$ degree rotation of one of the
Chern-Simons layers can be verified in experiments with neutral
atoms having QED dipole interaction with an electromagnetic field.
\end{abstract}

\maketitle

\section{Introduction}\label{sec1}
The Casimir effect \cite{CasPol, Casimir} is a quantum effect which
studies interaction between macroscopic objects in their ground
state. Interaction between two dielectric half-spaces separated by a
vacuum slit is determined by the Lifshitz formula \cite{Lifshitz}.
Theoretical study of the Casimir effect has received new
possibilities in the framework of scattering approach, the formalism
has been effectively applied to non-flat geometries including
diffraction gratings \cite{Mar2, Mar3, Marachevsky2020}, spheres and
cylinders \cite{ Emig0, Emig1, Jaffe1, Astrid1, Bordag1}. One can
find details of theoretical and experimental research in various
reviews and books on the subject \cite{MarSid, LP, Ginzburg1,
Ginzburg2, Plunien, Rep1, Santangelo3, Miltonreview, Jaffe3,
BuhmannScheel, K2, micro,Marachevskyreview, Woods3, Woods4, Lu1,
rev10, rev11, Vassilevich3, Buhmann3, Casbook}.

Chern-Simons action modifies the Casimir interaction essentially,
its study within 2+1 Abelian electrodynamics with Chern-Simons term
has been started in Ref. \cite{MiltonNg} where the
Maxwell-Chern-Simons electrodynamics has a massive spin-1
excitation. Chern-Simons constants of the layers are dimensionless
in 3+1 case. Rigid nonpenetrable boundary conditions modified by a
Chern-Simons term in 3+1 case have been considered in Refs.
\cite{Vascond1, Vascond2}, the Hall conductivity is not described by
these conditions. The Casimir energy of  two flat Chern-Simons
layers in vacuum has been derived in Refs. \cite{Pismak1, Mar20},
Casimir attraction and repulsion due to Chern-Simons boundary layers
on dielectric and metal half-spaces have been studied in Refs.
\cite{Mar21, Mar22}.

The Casimir-Polder potential for an anisotropic atom is obtained by
direct application of quantum electrodynamics in the second order
perturbation theory \cite{Mar40, Paulo1, Mar41, Johnson1, Mar42}.
The Casimir-Polder effect for  conducting planes has been considered
in Refs.\cite{Woods1, Woods2}, the Casimir-Polder effect for
conducting planes with a tensorial conductivity
\cite{Marachevskygraphene} has been considered in Refs.\cite{MarSid,
graphenereview, Nail1}. The Casimir-Polder potential of a neutral
anisotropic atom in the presence of a plane Chern-Simons layer has
been derived in Ref. \cite{Chern}, charge-parity violating effects
due to Chern-Simons layer have been investigated in Ref.
\cite{VMStefan}.

In the low-energy effective theory of topological insulators there
is a term proportional to $\theta \vec{E}\vec{H}$ in addition to the
standard electromagnetic energy density, this action can be
integrated over the volume of topological insulator into
Chern-Simons action at the boundary. The parameter $a$ of
Chern-Simons action is quantized in this case as follows: $a= \alpha
\theta/(2\pi)$, $\theta= (2m+1) \pi$, $\alpha$ is QED fine structure
constant, $m$ is an integer number \cite{top1}. Various aspects of
the Casimir interaction of topological insulators have been studied
in literature \cite{top2, top3, top4, top5, top6, top7}.

Theoretical description of Chern insulators \cite{CI1, CI2, CI3} is
given in a non-dispersive case by Chern-Simons action with the
parameter $a = C \alpha$, $C$ is a Chern number - topological
invariant giving the winding number of a map from a two-dimensional
torus to a two-dimensional unit sphere.  The Casimir interaction of
 Chern insulators is studied in Refs.
\cite{Mar20, top8, top9}.

Quantum Hall layers in external magnetic field also lead to a
quantized Casimir force, the parameter of the Chern-Simons action
$a= \nu \alpha$, $\nu$ is an integer or a fractional number
characterizing the plateau of the quantum Hall effect \cite{Mar21,
Mac, QHE1}.

The Casimir repulsion attracts a special attention in the Casimir
effect research, repulsion is a promising regime from the point of
view of technology. Rotation of polarization after reflection of the
electromagnetic wave from the Chern-Simons plane layer is an
important property which leads to regimes of attraction and
repulsion in the Casimir pressure between two Chern-Simons plane
parallel layers in vacuum and on boundaries of dielectrics or metals
\cite{Pismak1, Mar20, Mar21, Mar22}. Repulsive Casimir pressure has
not been investigated experimentally in this geometry so far.

A complementary way to study the Casimir effect is a local probe of
vacuum by a neutral atom in its ground state. It is tempting to
study vacuum between two Chern-Simons plane parallel layers locally
due to intriguing properties of this system.
This paper fills the gap in an important direction of local study of
vacuum in geometry of two Chern-Simons layers. Analytic results for
the Casimir-Polder potential of an anisotropic atom between two
Chern-Simons plane parallel layers in vacuum and on boundaries of
dielectric half-spaces are derived for the first time in the present
work.

Recently the formalism based on Green functions scattering has been
introduced \cite{MarSid}; in this approach one evaluates electric,
magnetic Green functions and the Casimir pressure in an explicit
gauge-invariant derivation.
 In Ref.\cite{MarSid} we have
derived the Casimir pressure and the Casimir-Polder potential in
systems without rotation of polarizations after reflection of
electromagnetic waves from boundaries between different media.

In the present paper we develop a principal generalization of the
Green functions scattering approach to a general case of reflection
from plane boundaries. In the presence of several Chern-Simons
layers one can not express the Casimir-Polder potential in terms of
two reflection coefficients (for TE and TM modes) even for a
diagonal tensor of atomic polarizability due to rotation of TE and
TM polarizations after reflection of the electromagnetic field from
each Chern-Simons layer. The matrix of reflection coefficients is
non-diagonal in this case \cite{Mar20, Mar21}. Derivation of the
Casimir-Polder potential in the presence of several Chern-Simons
layers has required a development of a novel technique presented in
this paper. We derive new formulas for the Casimir-Polder potentials
for all systems considered in this paper. We also discover and
investigate novel three-body vacuum effects in atom - two layers
system due to $180$ degree rotation of one of the layers.

We proceed as follows. In Section II we write expressions for the
field of a point dipole in vacuum in terms of electric and magnetic
fields following Ref.\cite{MarSid} and generalize Green functions
scattering formalism to the important case of non-diagonal
reflection matrices. Then we derive the result for the
Casimir-Polder potential of an anisotropic atom in the presence of a
 Chern-Simons plane boundary layer on a dielectric half-space. In Section
III we derive a general result for the Casimir-Polder potential of
an anisotropic atom between two dielectric half-spaces with
Chern-Simons plane parallel boundary layers. In Section IV we derive
results for the Casimir-Polder potential of an anisotropic atom
between two Chern-Simons plane parallel layers in vacuum expressed
through Lerch transcendent functions and polylogarithms. Section V
is devoted to the analysis of P-odd three-body vacuum effects,
experiments to measure the Casimir-Polder potential in the slit are
outlined.

Magnetic permeability of materials $\mu=1$ throughout the text. We
use $\hbar=c=1$ and Heaviside-Lorentz units.

\section{The Casimir-Polder Potential of an Anisotropic Atom above a Dielectric Half-Space
with Chern-Simons boundary layer} \label{sec2}



Green functions scattering method has been introduced in
Ref.\cite{MarSid} where it has been applied to derivation of various
classical results for the Casimir-Polder potential and the Casimir
pressure in geometries with plane boundaries, an explicit
gauge-invariant derivation of results has been worked out. All
results in Ref.\cite{MarSid} are expressed in terms of reflection
coefficients for TE and TM modes for problems when no mixing of TE
and TM modes is present after reflection of the electromagnetic wave
from the plane boundary between different media.

 Chern-Simons boundary layer
rotates each polarization of the incoming electromagnetic field
after reflection from the layer, rotation of polarizations is
described in this case by a non-diagonal reflection matrix
\cite{Mar20, Mar21}. Green functions scattering method is
generalized in this work to a general non-diagonal reflection
problem when applied to derivation of the Casimir-Polder potential.
The generalized formalism is developed and presented in detail in
this paper.

The result for the Casimir-Polder potential of a neutral anisotropic
atom interacting with Chern-Simons plane layer in vacuum is derived
in Ref.\cite{Chern}. In this Section we generalize the result of
Ref.\cite{Chern} and derive the Casimir-Polder potential of a
neutral anisotropic atom in its ground state located at a distance
$z_0$ from a dielectric half-space with a plane Chern-Simons
boundary layer.

Consider a dipole source at the point $\mathbf{r^\prime} = (0, 0,
z_0)$ characterized by electric dipole moment $d^l(t)$ with
components of the four-current density \cite{Chern}
\begin{align}
\rho (t, \mathbf{r}) &= -d^l(t)\partial_l\delta^3(\mathbf{r-r^\prime}) \: , \label{4density1} \\
 j^l (t, \mathbf{r}) &=  \partial_t d^l(t)\delta^3(\mathbf{r-r^\prime}) \: . \label{4current1}
\end{align}

Exact electric Green function can be found from the electric field
part solution of Maxwell equations for the electromagnetic field
propagating from a dipole source
(\ref{4density1}),(\ref{4current1}). The scattered electric Green
function $D^{E}_{ij}(t_1-t_2,\mathbf{r},\mathbf{r^{\prime}})$ is a
difference of the exact electric Green function and the vacuum
electric Green function. The Casimir-Polder potential is defined in
terms of the scattered electric Green function
$D^{E}_{ij}(t_1-t_2,\mathbf{r},\mathbf{r^{\prime}})$ from the source
(\ref{4density1}),(\ref{4current1}) and the atomic polarizability
$\alpha_{ij}(t_1 - t_2)= i\langle T(\hat{d}_i(t_1),
\hat{d}_j(t_2))\rangle $ as follows \cite{MarSid}:
\begin{equation}
U(z_0)= - \int\limits_0^{\infty} \frac{d\omega}{2\pi}
\alpha^{ij}(i\omega) D^{E}_{ij} (i\omega, \mathbf{r^{\prime}},
\mathbf{r^{\prime}}) . \label{CPpotential1}
\end{equation}

From Weyl formula \cite{Weyl}
\begin{equation}
\frac{e^{i\omega|\mathbf{r^{\prime}-r}|}}{4\pi
|\mathbf{r^{\prime}-r}| }= i \iint
\frac{e^{i(k_{x}(x^\prime-x)+k_{y}(y^\prime-y)+\sqrt{\omega^2-k^2_{x}-k^2_{y}}(z^\prime
-z))}} {2\sqrt{\omega^2-k^2_{x}-k^2_{y}}}
\frac{dk_{x}dk_{y}}{(2\pi)^2} \: ,
\end{equation}
valid for $z^{\prime} - z >0$, one can write electric and magnetic
fields propagating downwards from the dipole source
(\ref{4density1}),(\ref{4current1}) in the form \cite{MarSid}
\begin{align}
\mathbf{E^{0}}(\omega, \mathbf{r}) &= \int
\widetilde{\mathbf{N}}(\omega,\mathbf{k_\parallel} )
e^{i\mathbf{k_{\parallel}
\cdot r_{\parallel}}}e^{-ik_{z}(z-z_0)}d^2\mathbf{k_{\parallel}} , \label{electricsc}\\
\mathbf{H^{0}}(\omega, \mathbf{r}) &= \frac{1}{\omega}\int
[\widetilde{\mathbf{k}} \times \widetilde{\mathbf{N}}
(\omega,\mathbf{k_\parallel} )] e^{i\mathbf{k_{\parallel}\cdot
r_{\parallel}}}e^{-ik_{z}(z-z_0)} d^2\mathbf{k_{\parallel} } ,
\label{magneticsc} \\
\widetilde{\mathbf{N}}(\omega,\mathbf{k_\parallel} ) &=
\frac{i}{8\pi^2
  k_z}\left( -(\mathbf{d} \cdot \widetilde{\mathbf{k}})\mathbf{\widetilde{k}}+
\omega^2\mathbf{d}\right) , \label{Ntilde}
\end{align}
where $\mathbf{k_{\parallel}} = (k_x, k_y) $, $k_z= \sqrt{\omega^2-
k_{\parallel}^2}$, $\mathbf{\widetilde{k}}=(\mathbf{k_{\parallel}} ,
-k_z)$.  Components of the vacuum electric Green function for
$z^{\prime}-z>0$ can be determined from (\ref{electricsc}).

Consider a diffraction problem on a homogeneous dielectric
half-space $z < 0$ characterized by a dielectric permittivity
$\varepsilon(\omega)$ and a plane Chern-Simons boundary layer at
$z=0$ described by the action
\begin{equation}
S_{CS} = \frac{a}{2} \int \varepsilon^{z\nu\rho\sigma} A_{\nu}
F_{\rho\sigma} \,\, dt dx dy .
\end{equation}

To solve a diffraction problem we write electric and magnetic fields
for $z>0$ in the form
\begin{align}
\mathbf{E^1}(\omega, \mathbf{r}) &= \int
\widetilde{\mathbf{N}}(\omega,\mathbf{k_\parallel} )
e^{i\mathbf{k_{\parallel} \cdot
r_{\parallel}}}e^{-ik_{z}(z-z_0)}d^2\mathbf{k_{\parallel}}
+\int\mathbf{v}(\omega,\mathbf{k_\parallel} )
e^{i\mathbf{k_{\parallel}\cdot
r_{\parallel}}}e^{ik_{z}z} d^2\mathbf{k_{\parallel}} \: , \label{electricsc2}\\
\mathbf{H^1}(\omega, \mathbf{r}) &= \frac{1}{\omega}\int
[\widetilde{\mathbf{k}} \times \widetilde{\mathbf{N}}
(\omega,\mathbf{k_\parallel} )] e^{i\mathbf{k_{\parallel}\cdot
r_{\parallel}}}e^{-ik_{z}(z-z_0)} d^2\mathbf{k_{\parallel} }
 \nonumber \\
&+\frac{1}{\omega}\int [\mathbf{k \times
v}(\omega,\mathbf{k_\parallel)}] e^{i\mathbf{k_{\parallel} \cdot
r_{\parallel}}}e^{ik_{z}z} d^2\mathbf{k_{\parallel}} \: ,
\label{magneticsc2}
\end{align}
and for $z<0$ in the form
\begin{align}
\mathbf{E^2}(\omega, \mathbf{r}) &=
\int\mathbf{u}(\omega,\mathbf{k_\parallel} )
e^{i\mathbf{k_{\parallel}\cdot
r_{\parallel}}}e^{-iK_{z}z} d^2\mathbf{k_{\parallel}} \: , \\
\mathbf{H^2}(\omega, \mathbf{r}) &=
\frac{1}{\omega}\int\left(\mathbf{[k_{\parallel}\times u}(\omega,
\mathbf{k_\parallel)]}-K_z\mathbf{[n\times
u}(\omega,\mathbf{k_\parallel
)]}\right)e^{i\mathbf{k_{\parallel}\cdot r_{\parallel}}}e^{-iK_{z}z}
d^2\mathbf{k_{\parallel}} \:
\end{align}
with $K_z =\sqrt{\varepsilon(\omega)\omega^2 - k_x^2 - k_y^2}$ and $
\mathbf{n} = (0,0,1)$. Unknown vector functions
$\mathbf{v}(\omega,\mathbf{k_\parallel} )$ and
$\mathbf{u}(\omega,\mathbf{k_\parallel} )$ can be found from the
system of boundary conditions imposed on electric and magnetic
fields:
\begin{equation}
 {\rm div} (\mathbf{E^1} - \mathbf{E^0})=0,
 \end{equation}
 \begin{equation}
{\rm div} \mathbf{E^2}=0,
\end{equation}
\begin{equation}
E_x^1|_{z=0}=E_x^2|_{z=0},
\end{equation}
\begin{equation}
 E_y^1|_{z=0}=E_y^2|_{z=0},
 \end{equation}
 \begin{equation}
H_x^1|_{z=0+}-H_x^2|_{z=0-} = 2 a E_x^1|_{z=0},  \label{B1}
\end{equation}
\begin{equation}
H_y^1|_{z=0+}-H_y^2|_{z=0-} = 2 a E_y^1|_{z=0}.  \label{B2}
 \end{equation}
Boundary conditions (\ref{B1}), (\ref{B2}) have been considered in a
study of propagation of a plane electromagnetic wave in a medium
with a piecewise constant axion field \cite{BC1} and in a medium
with Chern-Simons layers \cite{BC2}. Note that the parameter $a$ is
proportional to a non-diagonal part of the surface conductivity
\cite{top7}. With this understanding the frequency dispersion $
a(\omega)$ may be considered in boundary conditions (\ref{B1}),
(\ref{B2}). To simplify notations we do not write explicitly the
frequency $\omega$ in $a(\omega)$ in what follows. In the
Casimir-Polder potential formulas we implicitly assume $a(i\omega)$
dependence.

It is convenient to use polar coordinates in two dimensional $(k_x,
k_y)$ momentum space and local orthogonal basis $\mathbf{e}_r,
\mathbf{e}_{\theta}, \mathbf{e}_z$ so that $\mathbf{k_{\parallel}}=
k_r \mathbf{e}_r$, $k_r = |\mathbf{k_{\parallel}} | $. We write
boundary conditions in this basis
\begin{equation}
v_{r}k_r+k_{z}v_{z}=0,
\end{equation}
\begin{equation}
u_{r}k_r-K_{z}u_{z}=0,
\end{equation}
\begin{equation}
u_{r}=v_{r}+\widetilde{N}_re^{ik_z z_0},
\end{equation}
\begin{equation}
u_{\theta}=v_{\theta}+\widetilde{N}_\theta e^{ik_z z_0},
\end{equation}
\begin{equation}
-k_zv_{\theta}+k_z\widetilde{N}_\theta e^{ik_z
z_0}-K_{z}u_{\theta}=2 \omega a u_{r},
\end{equation}
\begin{equation}
k_zv_{r}-k_rv_{z}-k_z\widetilde{N}_re^{ik_z
z_0}-k_r\widetilde{N}_ze^{ik_z z_0} +K_{z}u_{r} + k_r u_{z} = 2
\omega a u_{\theta}
\end{equation}
and get
\begin{align}
v_{r}&=\biggl[-\frac{r_{TM}+a^2T}{1+a^2T}\widetilde{N}_r +
\frac{k_z}{\omega}\frac{a T}{1+a^2
T}\widetilde{N}_\theta\biggr]e^{ik_z z_0}, \label{vr} \\
v_{\theta}&=\biggl[-\frac{\omega}{k_z} \frac{a T}{1+a^2
T}\widetilde{N}_r + \frac{r_{TE}-a^2T}{1+a^2T}\widetilde{N}_\theta
\biggr]e^{ik_z z_0}, \label{vtheta} \\
v_{z}&=\frac{k_r}{k_z}\biggl[\frac{r_{TM}+a^2T}{1+a^2T}\widetilde{N}_r-
\frac{k_z}{\omega}\frac{a T}{1+a^2
T}\widetilde{N}_\theta\biggr]e^{ik_z z_0}, \label{vz}
\end{align}
where $r_{TM}$, $r_{TE}$ are Fresnel reflection coefficients
\begin{equation}
r_{TM}(\omega, k_r) = \frac{\varepsilon(\omega) k_z -
K_z}{\varepsilon(\omega) k_z + K_z} \: ,\:\:\:\:\:\: r_{TE}(\omega,
k_r) = \frac{k_z - K_z}{k_z + K_z}
\end{equation}
and
\begin{equation}
T(\omega, k_r) = \frac{4k_zK_{z}}{(k_z+K_{z})(\varepsilon(\omega)
k_z+K_{z})}.
\end{equation}
Note that we omit dependence of reflection and transmission
coefficients on $(\omega, k_r)$ in (\ref{vr})-(\ref{vz}) for
brevity.

At this point  it is convenient to define the local matrix $R$
resulting from equations (\ref{vr}), (\ref{vtheta}):
\begin{equation}
 R(a, \varepsilon(\omega), \omega, k_r) \equiv \frac{1}{1 + a^2 T}
\begin{pmatrix}

-r_{TM} - a^2 T & \frac{k_z}{\omega} a T \\
- \frac{\omega}{k_z} a T & r_{TE} - a^2 T

\end{pmatrix} . \label{Rmatrix}
\end{equation}

To find the reflected part of the electric field one should use
rotation between two local bases and make substitutions
\begin{align}
d_r &= d_x \cos\theta + d_y \sin\theta \: ,\\
d_\theta &= d_x \sin\theta - d_y \cos\theta  \: ,\\
v_{x} &= v_{r} \cos\theta + v_{\theta} \sin\theta \: ,\\
v_{y} &= v_{r} \sin\theta - v_{\theta}\cos\theta
\end{align}
for every given $\mathbf{k_{\parallel}}$ to the scattered field part
of the expression (\ref{electricsc2}) by use of (\ref{Ntilde}),
(\ref{vr}), (\ref{vtheta}), (\ref{vz}). In doing so and noting that
\begin{align}
\widetilde{N}_r &= \frac{i}{8\pi^2}(k_z
(d_x\cos\theta+d_y\sin\theta) +
k_r d_z), \\
\widetilde{N}_\theta &= \frac{i}{8\pi^2}\frac{\omega^2}{k_z}
(d_x\sin\theta-d_y\cos\theta) ,
\end{align}
we obtain local contributions to cartesian components of scattered
electric Green functions for coinciding arguments at the point of a
dipole source:
\begin{multline}
D^E_{xx} (\omega, k_r, \theta, z=z^{\prime}=z_0 ) =
\frac{i}{8\pi^2}\biggl[\biggl(R_{11}k_z\cos\theta +
R_{12}\frac{\omega^2}{k_z}\sin\theta\biggr)\cos\theta \\
+\biggl(R_{21}k_z\cos\theta +
R_{22}\frac{\omega^2}{k_z}\sin\theta\biggr)\sin\theta\biggr]e^{2ik_zz_0},
\label{Exx}
\end{multline}
\begin{multline}
D^E_{yy} (\omega, k_r, \theta, z=z^{\prime}=z_0 )
 =
 \frac{i}{8\pi^2}\biggl[\biggl(R_{11}k_z\sin\theta-R_{12}\frac{\omega^2}{k_z}\cos\theta\biggr)\sin\theta
\\
-\biggl(R_{21}k_z\sin\theta-R_{22}\frac{\omega^2}{k_z}\cos\theta\biggr)\cos\theta\biggr]e^{2ik_zz_0},
\label{Eyy}
\end{multline}
\begin{equation}
D^E_{zz} (\omega, k_r, \theta, z=z^{\prime}=z_0 ) =
-\frac{i}{8\pi^2}\frac{k_r^2}{k_z}R_{11}e^{2ik_zz_0}, \label{Ezz}
\end{equation}
\begin{multline}
D^E_{xy} (\omega, k_r, \theta, z=z^{\prime}=z_0 ) =
\frac{i}{8\pi^2}\biggl[\biggl(R_{11}k_z\sin\theta-R_{12}\frac{\omega^2}{k_z}\cos\theta\biggr)\cos\theta
 \\
+\biggl(R_{21}k_z\sin\theta-R_{22}\frac{\omega^2}{k_z}\cos\theta\biggr)\sin\theta\biggr]e^{2ik_zz_0},
\label{Exy}
\end{multline}
\begin{multline}
D^E_{yx} (\omega, k_r, \theta, z=z^{\prime}=z_0 ) =
\frac{i}{8\pi^2}\biggl[\biggl(R_{11}k_z\cos\theta+R_{12}\frac{\omega^2}{k_z}\sin\theta\biggr)\sin\theta
\\ - \biggl(R_{21}k_z\cos\theta +
R_{22}\frac{\omega^2}{k_z}\sin\theta\biggr
)\cos\theta\biggr]e^{2ik_zz_0}, \label{Eyx}
\end{multline}
\begin{align}
&D^E_{xz} (\omega, k_r, \theta, z=z^{\prime}=z_0 ) =
\frac{i}{8\pi^2}\biggl[R_{11}k_r\cos\theta+R_{21}k_r\sin\theta\biggr]e^{2ik_zz_0},
\label{Exz} \\
&D^E_{zx}(\omega, k_r, \theta, z=z^{\prime}=z_0 ) =
\frac{i}{8\pi^2}\biggl[-R_{11}k_r\cos\theta+R_{21}k_r\sin\theta\biggr]e^{2ik_zz_0},
\label{Ezx} \\
&D^E_{yz}(\omega, k_r, \theta, z=z^{\prime}=z_0 ) =
\frac{i}{8\pi^2}\biggl[R_{11}k_r\sin\theta-R_{21}k_r\cos\theta\biggr]e^{2ik_zz_0},
\label{Eyz} \\
&D^E_{zy} (\omega, k_r, \theta, z=z^{\prime}=z_0 ) =
\frac{i}{8\pi^2}\biggl[-R_{11}k_r\sin\theta-R_{21} k_r
\cos\theta\biggr]e^{2ik_zz_0}. \label{Ezy}
\end{align}

The Casimir-Polder potential of an anisotropic atom above a
dielectric half-space with a plane Chern-Simons boundary layer is
found by integrating expressions (\ref{Exx})-(\ref{Ezy}) over polar
coordinates and making use of the formula (\ref{CPpotential1}) (we
separately write contributions to the Casimir-Polder potential from
different components of $\alpha_{ij}(i\omega)$):
\begin{multline}
U_{xx}(z_0)+U_{yy}(z_0)=-\frac{1}{16\pi^2}\int\limits_0^\infty
d\omega \bigl(\alpha_{xx}(i\omega) + \alpha_{yy}(i\omega) \bigr)\\
\times \int\limits_0^\infty dk_r k_r e^{-2k_z z_0}
\biggl(\frac{r_{TM}+a^2 T}{1+a^2 T}k_z-\frac{r_{TE}-a^2
T}{1+a^2T}\frac{\omega^2}{k_z}\biggr) ,  \label{UUxx}
\end{multline}
\begin{align}
&U_{zz}(z_0)=-\frac{1}{8\pi^2}\int\limits_0^\infty d\omega
\alpha_{zz}(i\omega) \int\limits_0^\infty dk_r \frac{k_r^3}{k_z}
e^{-2k_z z_0}\frac{r_{TM}+a^2 T}{1+a^2 T} , \\
&U_{xy}(z_0)+U_{yx}(z_0) = -\frac{1}{8\pi^2}\int\limits_0^\infty
d\omega \,\omega \bigl(\alpha_{xy}(i\omega) -
\alpha_{yx}(i\omega)\bigr) \int\limits_0^\infty dk_r k_r e^{-2k_z
z_0} \frac{ a T}{1+a^2 T} ,
\end{align}
\begin{equation}
U_{xz}=U_{zx}=U_{yz}=U_{zy}=0. \label{UUxz}
\end{equation}
Note that the Casimir-Polder potential (\ref{UUxx})-(\ref{UUxz}) has
a contribution of an antisymmetric part of the atomic polarizability
\cite{Khriplovich}. For a plane Chern-Simons layer in vacuum the
result of Ref.\cite{Chern} can be deduced from the formulas
(\ref{UUxx})-(\ref{UUxz}).

\section{The Casimir-Polder Potential of an Anisotropic Atom between Two Dielectric
half-spaces with Chern-Simons boundary layers}\label{sec3}

Geometry of two Chern-Simons plane parallel layers in vacuum or on
boundaries of dielectrics is of particular interest due to
prediction of repulsive and attractive Casimir pressure regimes
\cite{Pismak1, Mar20, Mar21, Mar22}. For two Chern-Simons plane
parallel layers in vacuum and the condition $a_1=a_2$ the Casimir
repulsion holds in an interval $a_1 \in [0, a_0]$, where $a_0
\approx 1.032502$ \cite{Pismak1,Mar21}, while for $a_1=-a_2$ the
Casimir attraction holds for all values of the parameter $a_1$
\cite{Mar20}.

It is definitely important to probe analogous geometry locally by
inserting neutral atoms into a cavity with Chern-Simons boundary
layers. The Casimir-Polder potential determines quantum interaction
of an anisotropic neutral atom in its ground state with cavity
walls, it depends on geometry and material of the cavity. Local
probe of the cavity with parallel plane boundaries by neutral atoms
is really promising from the experimental point of view since in
this case one avoids expected problems with parallelism in
measurements of the Casimir forces in geometries with parallel plane
boundaries.

\begin{figure}
\begin{center}
\includegraphics[width=130mm]{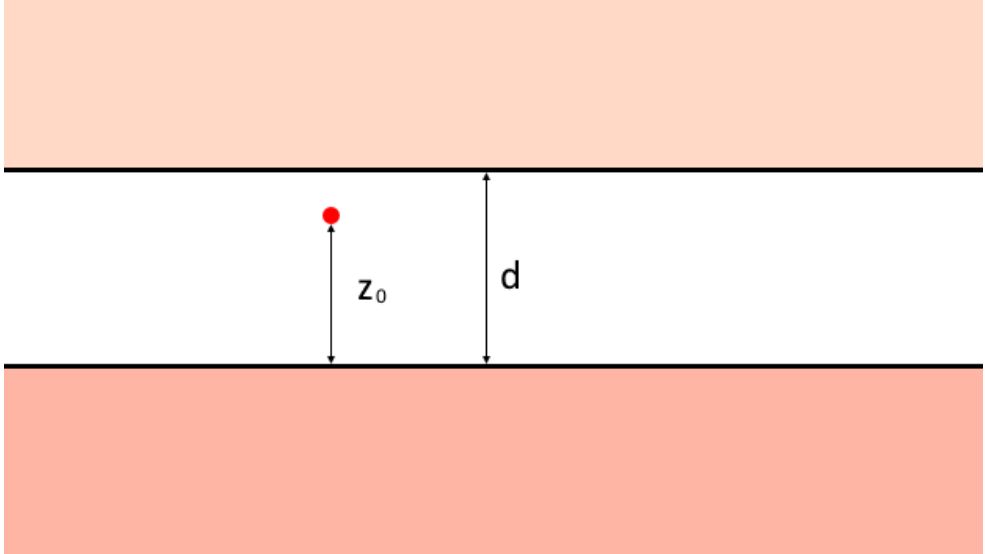}
\caption{Anisotropic neutral atom between two dielectric half-spaces
with plane Chern-Simons boundary layers, $z_0$ is a distance of the
atom from the layer and the dielectric medium characterized by the
index $2$, $d$ is a width of the vacuum slit.} \label{GT1}
\end{center}
\end{figure}

 Consider two
dielectric half-spaces  $z > d$, $z < 0$ with dielectric
permittivities $\varepsilon_1(\omega)$ and $\varepsilon_2(\omega)$
respectively and the vacuum slit $0 < z < d$ between them. Two
Chern-Simons plane parallel boundary layers are located at $z=d$ and
$z=0$ and characterized by the parameters $a_1(\omega)$ and
$a_2(\omega)$ respectively (see a discussion after (\ref{B2})). We
omit frequency dispersion in $a_1(\omega)$, $a_2(\omega)$ for
brevity in what follows as before. The atom is located at the point
$\mathbf{r^\prime} = (0, 0, z_0)$, $0 < z_0 < d$ (see
Fig.\ref{GT1}). In this Section we derive a general result for the
Casimir-Polder potential of a neutral anisotropic atom in this
system.

First it is convenient to solve a diffraction problem from an upper
half-space ($z \ge d$) when the lower half-space is absent. Consider
an upward propagation of an electromagnetic field from a point
dipole located at $\mathbf{r^\prime} = (0, 0, z_0)$, $z_0<d$. For
$z<d$ the expansions for electric and magnetic fields can be written
as follows:
\begin{align}
\mathbf{E^{1}}(\omega, \mathbf{r}) &= \int
\mathbf{N}e^{i\mathbf{k_{\parallel}\cdot r_{\parallel}}}
e^{ik_z(z-z_0)} d^2\mathbf{k_{\parallel}} +
\int \mathbf{v_1} e^{i\mathbf{k_{\parallel}\cdot r_{\parallel}}}e^{-ik_z z} d^2\mathbf{k_{\parallel}} \: , \label{expansion1} \\
\mathbf{H^{1}}(\omega, \mathbf{r}) &= \frac{1}{\omega}\int
[\mathbf{k}\times\mathbf{N}] e^{i\mathbf{k_{\parallel}\cdot
r_{\parallel}}} e^{ik_z(z-z_0)}
d^2\mathbf{k_{\parallel}}  \nonumber \\
&+\frac{1}{\omega}\int\left(\mathbf{[k_{\parallel}\times
v_1]}-k_z\mathbf{[n\times v_1]}\right)e^{i\mathbf{k_{\parallel}\cdot
r_{\parallel}}}e^{-ik_{z}z}d^2\mathbf{k_{\parallel}} , \\
\mathbf{N} &=  \frac{i}{8\pi^2 k_z} \bigl(-(\mathbf{k\cdot
d})\mathbf{k} + \omega^2 \mathbf{d}\bigr) . \label{expansion4}
\end{align}
 The vector function $\mathbf{v_1}$
depends on  $\omega, \mathbf{k_{\parallel}}$, $z_0$, $d$ and the
dipole moment $\mathbf{d}$. For $z>d$ we write transmitted fields in
the form
\begin{align}
\mathbf{E^{2}}(\omega, \mathbf{r}) &= \int  \mathbf{u_1}
e^{i\mathbf{k_{\parallel}\cdot r_{\parallel}}}e^{iK_{z1} z} d^2\mathbf{k_{\parallel}} \: , \\
\mathbf{H^{2}}(\omega, \mathbf{r}) &=
\frac{1}{\omega}\int\left(\mathbf{[k_{\parallel}\times
u_1]}+K_{z1}\mathbf{[n\times
u_1]}\right)e^{i\mathbf{k_{\parallel}\cdot
r_{\parallel}}}e^{iK_{z1}z}d^2\mathbf{k_{\parallel}} .
\label{expansion2}
\end{align}

Note that the parameter $a_1$ enters boundary conditions
\begin{equation}
H_x^2|_{z=d+}-H_x^1|_{z=d-} = 2 a_1 E_x^1|_{z=d},
\end{equation}
\begin{equation}
H_y^2|_{z=d+}-H_y^1|_{z=d-} = 2 a_1 E_y^1|_{z=d}.
 \end{equation}

In analogy to Section 2 we find
\begin{align}
v_{1r}&=\biggl[-\frac{r_{TM_1}+a_1^2T_1}{1+a_1^2T_1}N_r+
\frac{k_z}{\omega}\frac{a_1T_1}{1+a_1^2T_1}N_\theta\biggr]e^{ik_z(2d-z_0)},
\\
v_{1\theta}&=\biggl[ -\frac{\omega}{k_z}\frac{a_1T_1}{1+a_1^2T_1}N_r
+ \frac{r_{TE_1}-a_1^2T_1}{1+a_1^2T_1}N_\theta
\biggr]e^{ik_z(2d-z_0)}, \\
v_{1z}&=-\frac{k_r}{k_z}\biggl[\frac{r_{TM_1}+a_1^2T_1}{1+a_1^2T_1}N_r-
\frac{k_z}{\omega}\frac{a_1T_1}{1+a_1^2T_1}N_\theta\biggr]e^{ik_z(2d-z_0)},
\end{align}
where $r_{TM_1}$, $r_{TE_1}$, $T_1$ are written for a medium with a
dielectric permittivity $\varepsilon_1(\omega)$.

Now we turn to a solution of a diffraction problem when both
half-spaces are present. It is convenient to define from
(\ref{Rmatrix}) the matrices $R_1(\omega)$ and $R_2(\omega)$ for a
reflection of tangential components of the electric field from the
media above and below the point dipole respectively in a local basis
$\mathbf{e}_r, \mathbf{e}_{\theta}, \mathbf{e}_z$:
\begin{equation}
R_1(\omega) \equiv R(a_1, \varepsilon_1(\omega), \omega, k_r),  \:
\: R_2(\omega) \equiv R(a_2, \varepsilon_2(\omega ), \omega, k_r) ,
\end{equation}
here the medium for $z \le 0$ is denoted by the index $2$. Then the
tangential local components of the electric field in the interval $0
< z < d$ from the point dipole (\ref{4density1}),(\ref{4current1})
located at $(0, 0, z_0)$ are expressed in terms of matrices
$R_1(\omega)$, $R_2(\omega)$ as follows:
\begin{multline}
\begin{pmatrix}
E_r\\
E_\theta
\end{pmatrix}
= \frac{e^{ik_zz}}{I-R_2R_1e^{2ik_zd}}\biggl[R_2R_1
\begin{pmatrix}
N_r\\
N_\theta
\end{pmatrix}
e^{ik_z(2d-z_0)}+R_2
\begin{pmatrix}
\widetilde{N_r}\\
\widetilde{N_\theta}
\end{pmatrix}
e^{ik_zz_0}\biggr]\\+\frac{e^{ik_z(2d-z)}}{I-R_1R_2e^{2ik_zd}}\biggl[R_1R_2
\begin{pmatrix}
\widetilde{N_r}\\
\widetilde{N_\theta}
\end{pmatrix}
e^{ik_zz_0}+R_1
\begin{pmatrix}
N_r\\
N_\theta
\end{pmatrix}
e^{-ik_zz_0}\biggr] , \label{MR}
\end{multline}
in (\ref{MR}) the local components of the electric field are
obtained by a summation of multiple reflections from media with
indices $1$ and $2$.

It is convenient to define four matrices entering (\ref{MR}) after
Wick rotation:
\begin{align}
 M^1 &= \bigl(I-R_2(i\omega)R_1(i\omega)e^{-2k_zd}\bigr)^{-1} R_2(i\omega)R_1(i\omega), \label{M1}\\
 M^2 &= \bigl(I-R_2(i\omega)R_1(i\omega)e^{-2k_zd}\bigr)^{-1} R_2(i\omega), \\
M^3 &= \bigl(I-R_1(i\omega)R_2(i\omega)e^{-2k_zd}\bigr)^{-1} R_1(i\omega)R_2(i\omega), \\
M^4 &= \bigl(I-R_1(i\omega)R_2(i\omega)e^{-2k_zd}\bigr)^{-1}
R_1(i\omega) . \label{M4}
\end{align}
Components of scattered electric Green functions can be expressed in
terms of matrices (\ref{M1})-(\ref{M4}) following the scheme
explicitly presented in equations (\ref{Exx})-(\ref{Ezy}). After
integration over polar coordinates we express scattered electric
Green functions at imaginary frequencies for coinciding arguments $
\mathbf{r} = \mathbf{r^{\prime}}$ in terms of matrix elements of
matrices (\ref{M1})-(\ref{M4}):
\begin{multline}\label{D11}
D^E_{xx}(i\omega, \mathbf{r} = \mathbf{r^{\prime}}) =
D^E_{yy}(i\omega, \mathbf{r}= \mathbf{r^{\prime}}) =
-\frac{1}{8\pi}\int\limits_0^{\infty} dk_r k_r \\
\times\biggl[k_z(e^{-2k_zd}M_{11}^1+e^{-2k_zz_0}M_{11}^2+e^{-2k_zd}M_{11}^3+e^{-2k_z(d-z_0)}M_{11}^4)\\+\frac{\omega^2}{k_z}
(e^{-2k_zd}M_{22}^1+e^{-2k_zz_0}M_{22}^2+e^{-2k_zd}M_{22}^3+e^{-2k_z(d-z_0)}M_{22}^4)\biggr],
\end{multline}
\begin{multline}
 D^E_{zz}(i\omega, \mathbf{r} =
\mathbf{r^{\prime}}) = -\frac{1}{4\pi}\int\limits_0^{\infty} dk_r
\frac{k_r^3}{k_z} \\
\times\biggl[-e^{-2k_zd}M_{11}^1+e^{-2k_zz_0}M_{11}^2-e^{-2k_zd}M_{11}^3+e^{-2k_z(d-z_0)}M_{11}^4)\biggr],
\label{DEzz}
\end{multline}
\begin{multline}
D^E_{xy}(i\omega, \mathbf{r} = \mathbf{r^{\prime}}) = -
D^E_{yx}(i\omega, \mathbf{r} =
\mathbf{r^{\prime}}) = -\frac{1}{8\pi}\int\limits_0^{\infty} dk_r k_r \\
\times\biggl[-\frac{\omega^2}{k_z}
(e^{-2k_zd}M_{12}^1+e^{-2k_zz_0}M_{12}^2+e^{-2k_zd}M_{12}^3+e^{-2k_z(d-z_0)}M_{12}^4)\\+k_z(e^{-2k_zd}M_{21}^1+e^{-2k_zz_0}M_{21}^2+e^{-2k_zd}M_{21}^3+e^{-2k_z(d-z_0)}M_{21}^4)\biggr],
\label{D14}
\end{multline}
\begin{equation}
\label{D2} D^E_{xz}(i\omega, \mathbf{r} =
\mathbf{r^{\prime}})=D^E_{zx}(i\omega, \mathbf{r} =
\mathbf{r^{\prime}})=D^E_{yz}(i\omega, \mathbf{r} =
\mathbf{r^{\prime}})=D^E_{zy}(i\omega, \mathbf{r} =
\mathbf{r^{\prime}})=0.
\end{equation}

Now one can substitute expressions (\ref{D11})-(\ref{D2}) into the
formula (\ref{CPpotential1}) and evaluate the Casimir-Polder
potential of an anisotropic atom between two dielectric half-spaces
with Chern-Simons plane parallel boundary layers. The Casimir-Polder
potential in the limit $a_1$, $a_2 \to \infty$ is derived in
Appendix A.

\section{The Casimir-Polder Potential of an Anisotropic Atom between two
Chern-Simons layers in vacuum}\label{sec4}

In this Section we derive analytic results for the Casimir-Polder
potential of an anisotropic atom between two Chern-Simons plane
parallel layers in vacuum separated by a distance $d$, the atom is
positioned at the point $(0,0, z_0)$. The layer characterized by the
parameter $a_1$ is located at $z=d$, the layer characterized by the
parameter $a_2$ is located at $z=0$.

 In the system under consideration
$\varepsilon(\omega)=1$ for $z < 0$ and $z > d$. In this case the
matrices (\ref{M1})-(\ref{M4})  have the form
\begin{multline} \label{M11}
M^1 = M^3 = -\frac{1}{(1+a_1^2)(1+a_2^2)
\det[I-R_1R_2e^{-2k_zd}]}\\\times
\begin{pmatrix}
a_1a_2(1-a_1a_2(1-e^{-2k_zd})) & a_1a_2(a_1+a_2)\frac{k_z}{\omega}\\
-a_1a_2(a_1+a_2)\frac{\omega}{k_z} & a_1a_2(1-a_1a_2(1-e^{-2k_zd}))
\end{pmatrix} ,
\end{multline}
\begin{multline}
M^2= -\frac{1}{(1+a_1^2)(1+a_2^2) \det[I-R_1R_2e^{-2k_zd}]}\\\times
\begin{pmatrix}
a_2^2(1+a_1^2(1-e^{-2k_zd})) & -a_2(1+a_1^2+a_1a_2e^{-2k_zd})\frac{k_z}{\omega}\\
a_2(1+a_1^2+a_1a_2e^{-2k_zd})\frac{\omega}{k_z} &
a_2^2(1+a_1^2(1-e^{-2k_zd}))
\end{pmatrix} ,
\end{multline}
\begin{multline}
M^4= -\frac{1}{(1+a_1^2)(1+a_2^2) \det[I-R_1R_2e^{-2k_zd}]}\\\times
\begin{pmatrix}
a_1^2(1+a_2^2(1-e^{-2k_zd})) & -a_1(1+a_2^2+a_1a_2e^{-2k_zd})\frac{k_z}{\omega}\\
a_1(1+a_2^2+a_1a_2e^{-2k_zd})\frac{\omega}{k_z} &
a_1^2(1+a_2^2(1-e^{-2k_zd}))
\end{pmatrix} , \label{M14}
\end{multline}
where
\begin{multline}\label{Zn}
\frac{1}{(1+a_1^2)(1+a_2^2) \det[I-R_1R_2e^{-2k_zd}]}
\\ =\frac{1}{1+ a_1^2+ a_2^2 + 2a_1a_2e^{-2k_zd} + a_1^2a_2^2(1-e^{-2k_zd})^2} =
\frac{\gamma_1}{1+\beta_1y}+\frac{\gamma_2}{1+\beta_2y}
\end{multline}
with $y = \exp(-2k_zd)$, $A=a_1^2a_2^2$, $B=2(a_1a_2-a_1^2a_2^2)$,
$C=(1+a_1^2)(1+a_2^2)$, $y_{1,2}= \frac{-B\pm\sqrt{B^2-4AC}}{2A} =
(a_1a_2-1 \pm i (a_1+a_2))/(a_1a_2)$, $\beta_1=-1/y_1$,
$\beta_2=-1/y_2$, $\gamma_1= 1/(A y_1(y_2-y_1))$, $\gamma_2=1/(A
y_2(y_1-y_2))$.

Decomposition of the denominator  in (\ref{Zn}) into two terms leads
to an analytic result for the Casimir-Polder potential in terms of
Lerch transcendent functions. We change variables
\begin{equation}
\int\limits_0^\infty k_rdk_r f(k_z)=\int\limits_\omega^\infty
k_zdk_z f(k_z)
\end{equation}
and use the integral
\begin{equation}
 G_0(\chi, \beta, \omega) \equiv \int\limits_\omega^\infty
\frac{e^{-2k_z\chi}}{1+\beta
e^{-2k_zd}}dk_z=\frac{1}{2d}\int\limits_0^{e^{-2\omega
d}}\frac{y^{\frac{\chi}{d}-1}}{1+\beta y}dy
=\frac{e^{-2\omega\chi}}{2d} \Phi\Bigl(-\beta e^{-2\omega d}, 1,
\frac{\chi}{d} \Bigr),\label{G0}
\end{equation}
where  $\Phi(\alpha_1, \alpha_2, \alpha_3)$ is a Lerch transcendent
function. Derivatives over the parameter $\chi$ are defined as
follows:
\begin{equation}
G_1(\chi, \beta, \omega) \equiv \frac{1}{2}\frac{d}{d\chi}G_0(\chi,
\beta, \omega) = -\int\limits_\omega^\infty
k_z\frac{e^{-2k_z\chi}}{1+\beta e^{-2k_zd}}dk_z ,
\end{equation}
\begin{equation}
G_2 (\chi, \beta, \omega) \equiv
\frac{1}{4}\frac{d^2}{d\chi^2}G_0(\chi, \beta, \omega) =
\int\limits_\omega^\infty k_z^2\frac{e^{-2k_z\chi}}{1+\beta
e^{-2k_zd}}dk_z . \label{G2}
\end{equation}
The Casimir-Polder potential of an anisotropic atom between the two
layers is derived by making use of (\ref{CPpotential1}),
(\ref{D11})-(\ref{D14}), (\ref{M11})-(\ref{Zn}) and
(\ref{G0})-(\ref{G2}):
\begin{multline}
U_{xx}(z_0, d)+U_{yy}(z_0, d) =
\frac{1}{16\pi^2}\sum\limits_{i=1,2}\gamma_i\int\limits_0^\infty
d\omega  (\alpha_{xx}(i\omega)  +
\alpha_{yy}(i\omega) ) \\
\times \biggl[-2a_1^2a_2^2 G_2(2d, \beta_i, \omega)+ 2(a_1^2a_2^2 -
a_1a_2) G_2(d, \beta_i, \omega) - a_2^2(1+a_1^2) G_2(z_0, \beta_i,
\omega)  \\ + a_1^2a_2^2 G_2(z_0+d, \beta_i, \omega) -
a_1^2(1+a_2^2) G_2(d-z_0, \beta_i, \omega) + a_1^2a_2^2 G_2(2d-z_0,
\beta_i, \omega)
\\ +\omega^2\biggl(-2a_1^2a_2^2 G_0(2d, \beta_i, \omega)+ 2(a_1^2a_2^2 - a_1a_2)
G_0(d, \beta_i, \omega) - a_2^2(1+a_1^2) G_0(z_0, \beta_i, \omega)
\\ + a_1^2a_2^2 G_0(z_0+d, \beta_i, \omega) - a_1^2(1+a_2^2)
G_0(d-z_0, \beta_i, \omega) + a_1^2a_2^2 G_0(2d-z_0, \beta_i,
\omega)  \biggr)\biggr] , \label{Uxx}
\end{multline}

\begin{multline}
U_{zz}(z_0, d) =
\frac{1}{8\pi^2}\sum\limits_{i=1,2}\gamma_i\int\limits_0^\infty
d\omega \, \alpha_{zz}(i\omega) \\
\times \biggl[2a_1^2a_2^2 G_2(2d, \beta_i, \omega) - 2(a_1^2a_2^2 -
a_1a_2) G_2(d, \beta_i, \omega) - a_2^2(1+a_1^2) G_2(z_0, \beta_i,
\omega)  \\ + a_1^2a_2^2 G_2(z_0+d, \beta_i, \omega) -
a_1^2(1+a_2^2) G_2(d-z_0, \beta_i, \omega) + a_1^2a_2^2 G_2(2d-z_0,
\beta_i, \omega)
\\ +\omega^2\biggl(-2a_1^2a_2^2 G_0(2d, \beta_i, \omega)+ 2(a_1^2a_2^2 - a_1a_2)
G_0(d, \beta_i, \omega) + a_2^2(1+a_1^2) G_0(z_0, \beta_i, \omega)  \\
-a_1^2a_2^2 G_0(z_0+d, \beta_i, \omega) + a_1^2(1+a_2^2) G_0(d-z_0,
\beta_i, \omega) - a_1^2a_2^2 G_0(2d-z_0, \beta_i, \omega)
\biggr)\biggr] ,  \label{Uzz}
\end{multline}

\begin{multline}
U_{xy}(z_0, d) + U_{yx}(z_0, d) =
\frac{1}{8\pi^2}\sum\limits_{i=1,2}\gamma_i\int\limits_0^\infty
d\omega \, \omega \, (\alpha_{xy}(i\omega) - \alpha_{yx}(i\omega) )  \\
\times \biggl[-2a_1a_2(a_1+a_2) G_1(2d, \beta_i, \omega) +
a_2(1+a_1^2) G_1(z_0, \beta_i, \omega) + a_1a_2^2 G_1(z_0+d,
\beta_i, \omega) \\+ a_1(1+a_2^2) G_1(d-z_0, \beta_i, \omega) +
a_2a_1^2 G_1(2d-z_0, \beta_i, \omega) \biggr] . \label{Uxy}
\end{multline}

One can express components of the Casimir-Polder potential
(\ref{Uxx})-(\ref{Uxy}) in terms of Lerch transcendent functions due
to relations
\begin{equation}
G_1(\chi, \beta, \omega) = -\frac{e^{-2\omega\chi}}{4d^2}
\biggl(2\omega d \, \Phi\Bigl(-\beta e^{-2\omega d}, 1,
\frac{\chi}{d}\Bigr) + \Phi\Bigl(-\beta e^{-2\omega d}, 2,
\frac{\chi}{d}\Bigr) \biggr)  ,
\end{equation}
\begin{multline}
G_2(\chi, \beta, \omega) = \frac{e^{-2\omega\chi}}{4d^3}
\biggl(2\omega^2 d^2 \, \Phi\Bigl(-\beta e^{-2\omega d}, 1,
\frac{\chi}{d}\Bigr) + 2\omega d \Phi\Bigl(-\beta e^{-2\omega d}, 2,
\frac{\chi}{d}\Bigr)  \\ +\Phi\Bigl(-\beta e^{-2\omega d}, 3,
\frac{\chi}{d}\Bigr) \biggr) .
\end{multline}

At large distances of the atom from the layers $z_0$, $d-z_0 \gg
\lambda_0 \equiv 2\pi/\omega_0$, $\lambda_1 \equiv 2\pi/\omega_1$,
$\lambda_2 \equiv 2\pi/\omega_2$  ($\lambda_0$ is a wavelength
corresponding to a typical absorption frequency of the atom
$\omega_0$, $\lambda_1$ and $\lambda_2$ are wavelengths of the
layers corresponding to absorption frequencies $\omega_1$,
$\omega_2$ of the layers) the Casimir-Polder potential can be
derived analytically for arbitrary values of constants $a_1$, $a_2$
($a_1=a_1(0)$ and $a_2=a_2(0)$ for $z_0$, $d-z_0 \gg \lambda_1$,
$\lambda_2$). Noting that
\begin{equation}
\int\limits_0^{\infty} d\omega \, G_2(\chi, \beta_i, \omega) = 3
\int\limits_0^{\infty} d\omega \,\omega^2 G_0(\chi, \beta_i, \omega)
= \frac{3}{8 d^4} \Phi\Bigl(y_i^{-1}, 4, \frac{\chi}{d}\Bigr) ,
\end{equation}
we find from (\ref{Uxx}),(\ref{Uzz}) the Casimir-Polder potential of
the atom between two Chern-Simons plane parallel layers at large
distances from the layers resulting from the symmetric part of the
atomic polarizability:
\begin{multline}
U_{s}(z_0,d)=  U_{s1}(z_0, d) + U_{s2}(d) = \frac{\alpha_{xx}(0) +
\alpha_{yy}(0) + \alpha_{zz}(0)}{32 \pi^2 d^4} \\ \times
\sum\limits_{i=1,2}\gamma_i \biggl[
-a_2^2(1+a_1^2)\Phi\Bigl(y_i^{-1}, 4, \frac{z_0}{d}\Bigr)
-a_1^2(1+a_2^2)\Phi\Bigl(y_i^{-1}, 4, \frac{d-z_0}{d}\Bigr)  \\
+a_1^2a_2^2 \Phi\Bigl(y_i^{-1}, 4, \frac{d+z_0}{d}\Bigr) +
a_1^2a_2^2 \Phi\Bigl(y_i^{-1}, 4, \frac{2d-z_0}{d}\Bigr) \biggr] +
U_{s2}(d) , \label{ld1}
\end{multline}
\begin{multline}
U_{s2}(d) = \frac{\alpha_{xx}(0) + \alpha_{yy}(0) -
\alpha_{zz}(0)}{32 \pi^2 d^4} \sum\limits_{i=1,2}  {\rm
Li}_4(y_i^{-1}) \\ = \frac{\alpha_{xx}(0) + \alpha_{yy}(0) -
\alpha_{zz}(0)}{32 \pi^2 d^4} \biggl({\rm
Li}_4\biggl(\frac{a_1a_2}{(a_1+i)(a_2+i)}\biggr) + {\rm
Li}_4\biggl(\frac{a_1a_2}{(a_1-i)(a_2-i)}\biggr) \biggr) ,
\end{multline}
${\rm Li}_4(z)$  is a polylogarithm function. For $a_2=-a_1$ one
finds from (\ref{ld1})
\begin{multline}
U_{s}(z_0, d) =  \frac{\alpha_{xx}(0) + \alpha_{yy}(0) +
\alpha_{zz}(0)}{32 \pi^2 d^4}  \\ \times
\biggl[ -\frac{a_1^2}{1+a_1^2} \biggl(\Phi_2\Bigl(\frac{a_1^2}{1+a_1^2}  , 4, \frac{z_0}{d}\Bigr) + \Phi_2\Bigl(\frac{a_1^2}{1+a_1^2} , 4, \frac{d-z_0}{d}\Bigr)\biggr) \\
+ \frac{a_1^4}{(1+a_1^2)^2} \biggl(\Phi_2\Bigl(\frac{a_1^2}{1+a_1^2}  , 4, \frac{d+z_0}{d}\Bigr) +
\Phi_2\Bigl(\frac{a_1^2}{1+a_1^2}  , 4, \frac{2d-z_0}{d}\Bigr)\biggr) \biggr]  + U_{s2}(d) ,  \label{opp}
\end{multline}
where
\begin{equation}
\Phi_2\Bigl(z, s, \alpha \Bigr) \equiv \Phi\Bigl(z, s, \alpha \Bigr)  + z \frac{\partial\Phi\Bigl(z, s, \alpha \Bigr)}{\partial z} .
\end{equation}

To find the leading contribution to the Casimir-Polder potential at
large distances from the layers resulting from the antisymmetric
part of the atomic polarizability tensor it is sufficient to take
into account the leading term in the expansion of the antisymmetric
part of the polarizability tensor for small $\omega$ \cite{Chern}:
$\alpha_{xy}(\omega) - \alpha_{yx}(\omega) \simeq i \omega C_{as}$.
In doing so, we obtain from (\ref{Uxy}) the leading contribution to
the Casimir-Polder potential of the atom between two Chern-Simons
plane parallel layers at large distances from the layers resulting
from the antisymmetric part of the atomic polarizability:
\begin{multline}
U_{as}(z_0, d)=\frac{C_{as}}{32\pi^2 d^5}\sum\limits_{i=1,2}\gamma_i
\biggl[ a_2(1+a_1^2)\Phi\Bigl(y_i^{-1}, 5, \frac{z_0}{d}\Bigr) +
a_1(1+a_2^2)\Phi\Bigl(y_i^{-1}, 5, \frac{d-z_0}{d}\Bigr)  \\
+a_1a_2^2 \Phi\Bigl(y_i^{-1}, 5, \frac{d+z_0}{d}\Bigr) + a_2a_1^2
\Phi\Bigl(y_i^{-1}, 5, \frac{2d-z_0}{d}\Bigr) \biggr] + U_{as2}(d) , \label{Uas1}
\end{multline}
\begin{equation}
U_{as2}(d) =  \frac{C_{as}}{16\pi^2 d^5} \Im\bigl(y_1{\rm Li}_5(y_1^{-1})\bigr) . \label{ld4}
\end{equation}
For $a_2=-a_1$ one finds from (\ref{Uas1}) and (\ref{ld4})
\begin{multline}
U_{as}(z_0, d) =    \frac{C_{as}}{32\pi^2
d^5} \biggl[ \frac{a_1}{1+a_1^2} \biggl(-\Phi_2\Bigl(\frac{a_1^2}{1+a_1^2}  , 5, \frac{z_0}{d}\Bigr) + \Phi_2\Bigl(\frac{a_1^2}{1+a_1^2} , 5, \frac{d-z_0}{d}\Bigr)\biggr) \\
+ \frac{a_1^3}{(1+a_1^2)^2} \biggl(\Phi_2\Bigl(\frac{a_1^2}{1+a_1^2}
, 5, \frac{d+z_0}{d}\Bigr) - \Phi_2\Bigl(\frac{a_1^2}{1+a_1^2}  , 5,
\frac{2d-z_0}{d}\Bigr)\biggr) \biggr]  . \label{ld5}
\end{multline}

In the limit $a_1$, $a_2 \to \infty$ the potential $U_{s}(z_0, d)$
is in agreement with Barton \cite{Barton} ($\rho=z_0/d$):
\begin{align}
U_{id}(z_0 , d) = -\frac{\pi^2}{96 d^4} &\bigl(\alpha_{xx}(0) +
\alpha_{yy}(0) + \alpha_{zz}(0)\bigr) \frac{3 - 2\sin^2(\pi
\rho)}{\sin^4(\pi \rho)} \nonumber
\\ +\frac{\pi^2}{1440 d^4} &\bigl(\alpha_{xx}(0) + \alpha_{yy}(0) -
\alpha_{zz}(0) \bigr) ,  \label{Barton1}
\end{align}
the asymptotics of $U_{s1}(z_0, d)$ at large $a_1$, $a_2$ is derived
in Appendix B.

\begin{figure}
\begin{center}
\includegraphics[width=130mm]{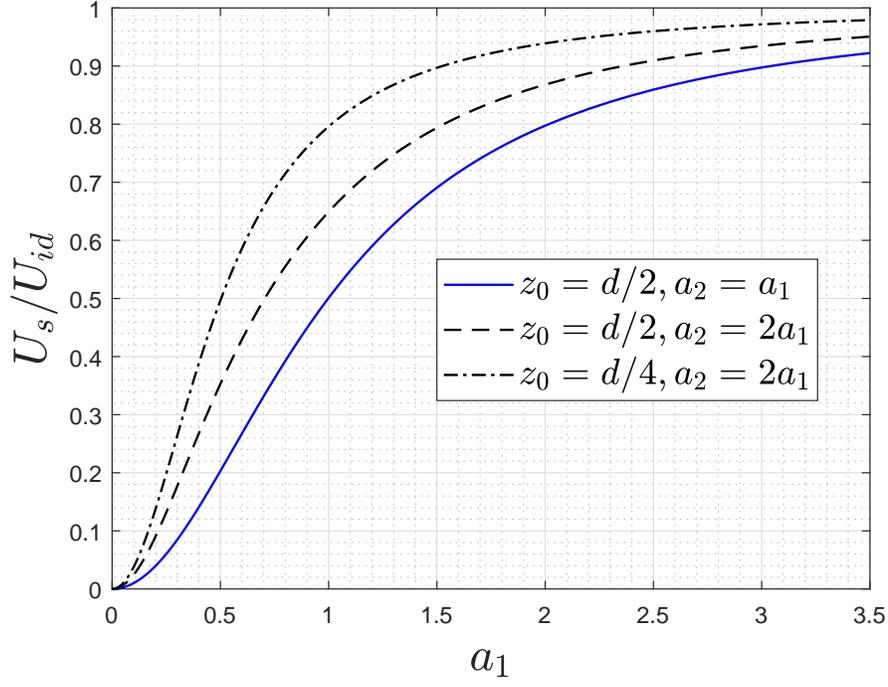}
 \caption{Ratios of the Casimir-Polder potential of a neutral
 polarizable isotropic atom located between two
plane Chern-Simons layers in vacuum $U_s(z_0, d)$ to the potential
of the same atom between two perfectly conducting planes
$U_{id}(z_0, d)$, here $z_0$ is a distance of the atom from the
layer characterized by a constant $a_2$, $d$ is a distance between
the layers.} \label{GT2}
\end{center}
\end{figure}

We use (\ref{ld1}) and (\ref{Barton1}) to evaluate the ratio of the
Casimir-Polder potential of a neutral polarizable atom in the
presence of two Chern-Simons plane parallel layers to the
Casimir-Polder potential of an atom in the presence of two perfectly
conducting parallel planes ($a_1 \to \infty$, $a_2 \to \infty$).
Ratios $U_s/U_{id}$ are shown in Fig.\ref{GT2} for an isotropic atom
with $ \alpha_{xx}(0) = \alpha_{yy}(0) = \alpha_{zz}(0)$ for
$a_1=a_2$ and $a_2=2a_1$ in an interval $a_1 \in [0, 3.5]$.

\section{P-odd vacuum effects}

Now we present the most intriguing result of the paper - theoretical
prediction of P-odd three-body vacuum effects. By P-odd three-body
effects we denote physical effects that differ after $180$ degree
rotation of one of the Chern-Simons layers in the presence of a
neutral atom. In our notations $180$ degree rotation of one of the
layers corresponds to the substitution $a_1 \to -a_1$ (or $a_2 \to
-a_2$) into the Casimir-Polder potential. Note that in the model
under consideration a neutral polarizable atom interacts via a
quantum electrodynamical dipole interaction with an electromagnetic
field.

\begin{figure}
\begin{center}
\includegraphics[width=130mm]{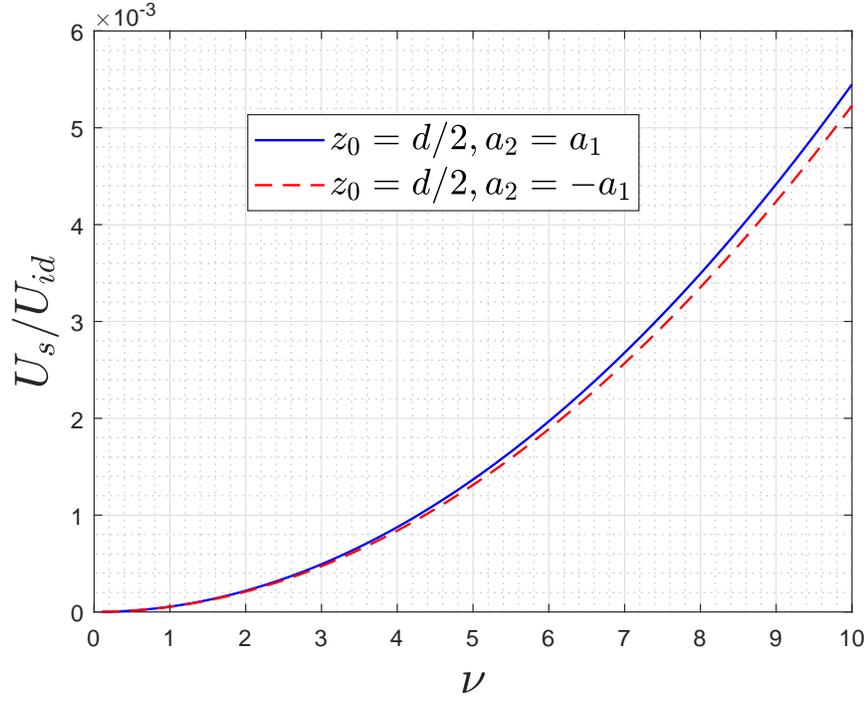}
\caption{Ratios of the Casimir-Polder potentials $U_s(z_0,
d)/U_{id}(z_0, d)$ differing by $180$ degree rotation of the
Chern-Simons layer characterized by a parameter $a_2$: $a_2=a_1$ and
$a_2=-a_1$. Here $z_0$ is a distance of the atom from the layer
characterized by a constant $a_2$, $d$ is a distance between the
layers, a dimensionless parameter $\nu= a_1/\alpha$ is quantized in
quantum Hall layers and Chern insulators.} \label{GT3}
\end{center}
\end{figure}

\begin{figure}
\begin{center}
\includegraphics[width=130mm]{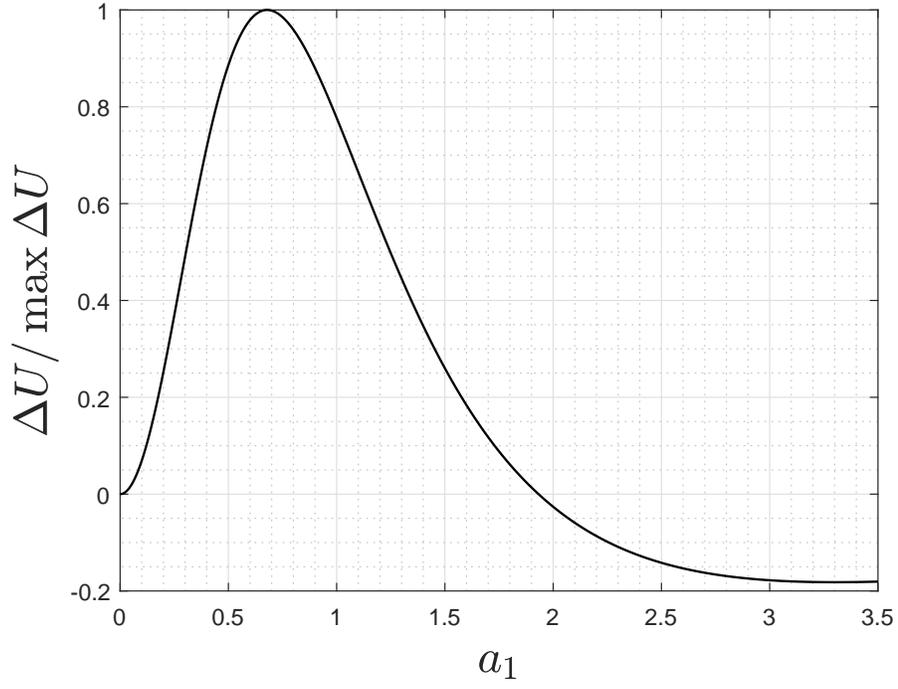}
 \caption{Ratio $\Delta U =
U_s (z_0=d/2, d, a_2=-a_1) - U_s (z_0=d/2, d, a_2=a_1)$ to
$\max\Delta U \approx 0.00587 |U_{id}(z_0=d/2, d)|$, $\max\Delta U$
holds at $a_1\approx 0.678$.} \label{GT4}
\end{center}
\end{figure}

From the formulas (\ref{ld1}), (\ref{opp}), (\ref{Barton1}) we find
ratios of potentials $U_s(z_0=d/2, d, a_2=a_1)$ and $U_s(z_0=d/2, d,
a_2=-a_1)$ to the potential $U_{id}(z_0=d/2, d)$ of the atom between
two perfectly conducting parallel planes and present these ratios
for $\nu= a_1/\alpha \le 10$ in Fig.\ref{GT3}. Note that the
parameter $\nu$ is quantized in quantum Hall layers and Chern
insulators. Values of analogous ratios for larger values of $\nu$
can be extracted from Fig.\ref{GT2} and Fig.\ref{GT4}.

In Fig.\ref{GT4} we compare the Casimir-Polder potentials of an
isotropic atom for two systems differing by parity of one of the
Chern-Simons layers: $a_2=a_1$ and $a_2=-a_1$. We use formulas
(\ref{ld1}), (\ref{opp}) to find ratio of the difference $\Delta U =
U_s (z_0=d/2, d, a_2=-a_1) - U_s (z_0=d/2, d, a_2=a_1)$ to
$\max\Delta U \approx 0.00587 |U_{id}(z_0=d/2, d)|$, $\max\Delta U$
holds at $a_1\approx 0.678$, the ratio is shown in Fig.\ref{GT4}. It
is interesting to note that the ratio $\max\Delta U/|U_{id}(z_0=d/2,
d)| \approx 0.00587$ for $a_1\approx 0.678$ ($\nu \approx 93$) is
even greater than ratios found in Fig.\ref{GT3} for $\nu = 10$.

Consider first a classical mechanics reasoning in a gedanken
experiment which demonstrates the way to study P-odd effects by
neutral atoms in the system of two Chern-Simons plane parallel
layers. Consider a neutral atom which starts moving in free space
from the point $A$ far away from the layers, continues its movement
between the layers so that $z_0=d/2$ and finally leaves the space
between the layers reaching the point $B$ in free space far away
from the layers ($A$, $z_0$ and $B$ are on the same straight line
parallel to the layers). The Casimir-Polder potential of the atom
between two Chern-Simons layers in this case is equal in absolute
value to an increase of the kinetic energy of the atom between the
layers. The atom moves at a higher speed between the layers than its
speed in vacuum, the time difference of the flights with and without
Chern-Simons layers can be measured in experiments. When one changes
the parameter $a_1$ (or $a_2$) by changing an external magnetic
field in case of a quantum Hall layer or by selecting a layer with a
different Chern number in case of a Chern insulator one changes
quantum vacuum and the value of the Casimir-Polder potential. At the
same time one changes the time of the flight of the atom from the
point $A$ to the point $B$. In summary, measuring timeshifts in
flight time of neutral atoms through the slit between two
Chern-Simons layers is a direct way to study energy shifts in the
Casimir-Polder potential due to changes in $a_1$, $a_2$.

Another possibility to study the Casimir-Polder potential is a
measurement of the number of atoms passing through a cavity. The
experiment \cite{Hinds} with sodium atoms passing through a
micron-sized cavity clearly proved existence of the Casimir-Polder
force  by measuring the intensity of a sodium atomic beam
transmitted through the cavity as a function of separation of cavity
boundaries. The experiment \cite{Hinds} can be considered as a
prototype of experiments for measurement of P-odd vacuum effects.

One can also study quantum effects of propagation of neutral atoms
in a slit between Chern-Simons layers in the presence of a
gravitational field. A combined effect of quantum reflection of
neutral atoms and the Earth's gravitational field during propagation
of atoms through the slit in analogy to experiments with neutrons
\cite{Nesvizh1, DM1} should expand experimental capabilities in
search of dark matter. Note that quantum reflection of atoms from
rigid boundaries arises due to an attractive rapidly changing
Casimir-Polder potential \cite{Astrid2}. Chern-Simons boundary
layers with P-odd vacuum effects lead to new opportunities in this
research direction.

\section{Conclusions}


In this paper  we develop a principal generalization of the Green
functions scattering method \cite{MarSid} for the case when one can
not express the Casimir-Polder potential in terms of diagonal
reflection matrix consisting of reflection coefficients for TE and
TM modes. Diffraction of an electromagnetic wave in a system with
Chern-Simons plane boundary layer is described by a non-diagonal
reflection matrix due to rotation of polarizations after reflection
of the incoming electromagnetic wave from the layer \cite{Mar20,
Mar21}. The technique developed in this paper is used to derive new
formulas for the Casimir-Polder potential of an anisotropic atom in
the presence of dielectric half-spaces with Chern-Simons plane
parallel boundary layers.

The technique developed in the present paper should be effective for
derivation of the Casimir-Polder potential of an anisotropic neutral
atom located between any media with plane parallel boundaries when
rotation of polarizations occurs after reflection from boundaries.
In general, once reflection of electric and magnetic fields from
plane parallel boundaries is defined, the Casimir-Polder potential
of an anisotropic neutral atom in the system can be found by
application of a technique developed in this work.


We have started from derivation of the Casimir-Polder potential of
an anisotropic atom in the presence of a dielectric half-space with
a Chern-Simons plane layer at its boundary, the  result is presented
in general formulas (\ref{UUxx})-(\ref{UUxz}). We have continued
with derivation of a general result for the Casimir-Polder potential
of an anisotropic atom between two dielectric half-spaces with
Chern-Simons plane parallel boundary layers, the result is given by
expressions (\ref{D11})-(\ref{D2}) when substituting into a
well-known formula (\ref{CPpotential1}). This general result is then
used to obtain formulas (\ref{Uxx})-(\ref{Uxy}) for the components
of the Casimir-Polder potential of an anisotropic atom between two
Chern-Simons plane parallel layers in vacuum expressed through Lerch
transcendent functions. The Casimir-Polder potential of the atom
between two Chern-Simons plane parallel layers at large distances of
the atom from both layers  is expressed through Lerch transcendent
functions and polylogarithms in formulas (\ref{ld1})-(\ref{ld5}).
All these results for the Casimir-Polder potentials are novel.

Knowledge of formulas for the Casimir-Polder potential of an
anisotropic atom between Chern-Simons layers in vacuum and on
dielectrics is important for precise comparison of the theory and
experiments discussed in Section V. Quantization of parameters
$a_1$, $a_2$ in topological insulators, Chern insulators and quantum
Hall layers leads to precise knowledge of the Casimir-Polder
potential of the atom at large separations from the boundaries of a
cavity with Chern-Simons boundary layers, which is relevant for
planning the experiments and conducting precise comparison of the
theory and experiments.

Novel P-odd effects for the Casimir-Polder potential between two
Chern-Simons plane parallel layers in vacuum due to a substitution
$a_2 \to -a_2$ are predicted and analyzed in Section V. P-odd
effects arise due to a three-body interaction between a neutral atom
in its ground state and two Chern-Simons layers. Our results
demonstrate that a neutral atom with QED dipole interaction may
become an effective tool for measurement of P-odd vacuum effects due
to $180$ degree rotation of one of the Chern-Simons layers.
Predicted dependence of the Casimir-Polder potential of a neutral
atom on $180$ degree rotation of one of the Chern-Simons layers in a
cavity suggests an intriguing fundamental experimental check of
quantum vacuum properties based on rotation of the topological
material.

\begin{acknowledgments}
This research has received financial support from the grant of
Russian Science Foundation (RSF project ${\rm N^{\underline{o}}}
\:\, 22-13-00151$ ). Research by V.N.M. and A.A.S. was performed at
the Research park of St.Petersburg State University Computing
Center.
\end{acknowledgments}


\appendix

\section{Perfectly conducting parallel planes}

In the limit $a_1, a_2 \to \infty$ we find from (\ref{Rmatrix}),
(\ref{M1})-(\ref{M4})
\begin{equation} M^1=M^3=-M^2=-M^4=
\frac{1}{1-e^{-2k_zd}}
\begin{pmatrix}
1 & 0 \\ 0 & 1
\end{pmatrix} . \label{Mperfect}
\end{equation}
For $a_1, a_2 \to \infty$ the Casimir-Polder potential of a neutral
atom with a frequency dispersion of the polarizability
 is derived from (\ref{CPpotential1}), (\ref{D11}),
(\ref{DEzz}), (\ref{Mperfect}):
\begin{multline}
U_{2}(z_0,d)= -\int\limits_0^\infty
\frac{d\omega}{2\pi}\int\limits_0^\infty \frac{dk_r k_r}{2\pi}
\frac{\exp({-2\sqrt{\omega^2+k_r^2}z_0}) +
\exp({-2\sqrt{\omega^2+k_r^2}(d-z_0)})}{4 \sqrt{\omega^2+k_r^2} (1- \exp({-2\sqrt{\omega^2+k_r^2}d}))}  \\
\times \Bigl[ (2\omega^2+k_r^2)
\bigl(\alpha_{xx}(i\omega)+\alpha_{yy}(i\omega)\bigr) + 2k_r^2
\alpha_{zz}(i\omega)  \Bigr]  \label{CPtwo} \\
 +\int\limits_0^\infty \frac{d\omega}{2\pi}\int\limits_0^\infty
\frac{dk_r k_r}{2\pi} \frac{\exp({-2\sqrt{\omega^2+k_r^2}d})}{2
\sqrt{\omega^2+k_r^2}(1-
\exp({-2\sqrt{\omega^2+k_r^2}d}))}  \\
\times \Bigl[
(2\omega^2+k_r^2)\bigl(\alpha_{xx}(i\omega)+\alpha_{yy}(i\omega)\bigr)
 -2 k_r^2 \alpha_{zz}(i\omega) \Bigr]  .
\end{multline}
The potential $U_{2}(z_0,d)$ (\ref{CPtwo}) coincides with the
Casimir-Polder potential of a neutral ani\-so\-tro\-pic atom between
two perfectly conducting parallel planes \cite{MarSid}.

\section{Asymptotics}

 At large distances of the atom from the
Chern-Simons layers the Casimir-Polder potential is derived in
(\ref{ld1}).  Now we consider the asymptotics of $U_{s1}(z_0, d)$ at
large $a_1$, $a_2$.

One can use equality $\gamma_1 + \gamma_2 = 1/((1+a_1^2)(1+a_2^2))$
and expansions $y_1^{-1}\sim y_2^{-1} \sim 1 - 1/a_1^2 - 1/a_2^2 -
1/(a_1a_2)$ to write
\begin{multline}
\sum\limits_{i=1,2}\gamma_i \Phi(y_i^{-1}, s, \alpha) =
\frac{\Phi(y_2^{-1}, s, \alpha) }{(1+a_1^2)(1+a_2^2)} +
\gamma_1 \bigl(\Phi(y_1^{-1}, s, \alpha) - \Phi (y_2^{-1}, s, \alpha)\bigr)  \\
\approx \frac{\Phi(y_2^{-1}, s, \alpha) +  \Phi^{\prime}(y_2^{-1},
s, \alpha) y_1^{-1} }{(1+a_1^2)(1+a_2^2)}  \approx
\frac{1}{(1+a_1^2)(1+a_2^2)} \biggl[ \Phi(1, s, \alpha) +
\Phi^{\prime}(1, s, \alpha)  \\ - \biggl(\frac{1}{a_1^2} +
\frac{1}{a_2^2} + \frac{1}{a_1a_2}\biggr) \bigl(2\Phi^{\prime}(1, s,
\alpha) + \Phi^{\prime\prime}(1, s, \alpha) \bigr)  \biggr]  ,
\label{expan1}
\end{multline}
derivatives in Lerch transcendent functions are taken by the first
argument. It is convenient to use integral representation of Lerch
transcendent function
\begin{equation}
\Phi(z, s, \alpha) = \frac{1}{\Gamma(s)}\int\limits_0^{\infty}
\frac{t^{s-1} e^{-\alpha t}}{1 - ze^{-t}} dt
\end{equation}
and expansion (\ref{expan1}) to express asymptotics of $U_{s1}(z_0,
d)$ in (\ref{ld1}) for large
 $a_1$, $a_2$  in terms of Hurwitz zeta function $\zeta(s, \alpha) = \Phi(1, s, \alpha)$ due to integral
\begin{equation}
\int\limits_0^{\infty} \frac{t^{s-1}e^{-\alpha t}}{(e^t -1)^2} dt =
\Gamma(s) [\zeta(s-1, \alpha+2) - (\alpha+1)\zeta(s, \alpha+2) ]
\end{equation}
as follows ($\rho = z_0/d$):
\begin{multline}
U_{s1}(z_0, d) \sim -\frac{\alpha_{xx}(0) + \alpha_{yy}(0) +
\alpha_{zz}(0)}{32 \pi^2 d^4} \biggl[ \biggl(1 - \frac{1}{a_1^2} -
\frac{1}{a_2^2}\biggr)
\Bigl(\zeta(4, \rho) + \zeta(4, 1- \rho)\Bigr)  \\
+\frac{1}{a_1^2}\Bigl(\zeta(3, \rho) + (1-\rho)\zeta(4, \rho)\Bigr)
+ \frac{1}{a_2^2}\Bigl(\zeta(3, 1-\rho) +\rho\,\zeta(4,
1-\rho)\Bigr)
 \\ - 2 \biggl(\frac{1}{a_1^2} + \frac{1}{a_2^2} +
\frac{1}{a_1a_2}\biggr) \Bigl(\zeta(3, -\rho+2) + \zeta(3, \rho+1)
\\ + (\rho-1)\zeta(4, 2-\rho) - \rho\,\zeta(4, \rho+1) \Bigr) \biggr]  .
\label{asympt1}
\end{multline}
Note that the asymptotics (\ref{asympt1}) contains the term
proportional to $1/(a_1a_2)$ which changes its sign during 180
degree rotation of one of the Chern-Simons layers ($a_1 \to -a_1$ or
$a_2 \to -a_2$). Note also that
\begin{equation}
\zeta(4, \rho) + \zeta(4, 1- \rho) = \frac{\pi^4}{3}\frac{3 -
2\sin^2(\pi \rho)}{\sin^4(\pi \rho)} ,
\end{equation}
so that in the limit $a_1 \to \infty$, $a_2 \to \infty$ one gets
(\ref{Barton1}).

\newpage

\newpage

\newpage

\newpage

\end{document}